\pdfoutput=1

\documentclass[aps,prl,notitlepage,twoside,twocolumn,preprintnumbers,showpacs,showkeys,byrevtex,floatfix,nofootinbib,amsmath,amssymb,]{revtex4-1}
\usepackage{amscd,array,dsfont,graphicx,mathtools,multirow,rotating,tensor,verbatim,youngtab}
\usepackage[bookmarksnumbered,bookmarksopen=true,bookmarksopenlevel=1,linktocpage,pdfstartview=FitH]{hyperref}
\usepackage[all]{hypcap}

\allowdisplaybreaks[1]

\newcolumntype{M}[1]{>{$}{#1}<{$}}

\newcolumntype{M}[1]{>{$}{#1}<{$}}

\newcommand{\sst}[1]{{\scriptscriptstyle #1}}

\def\0{{\sst{(0)}}}
\def\1{{\sst{(1)}}}
\def\2{{\sst{(2)}}}
\def\3{{\sst{(3)}}}
\def\4{{\sst{(4)}}}
\def\5{{\sst{(5)}}}
\def\6{{\sst{(6)}}}
\def\7{{\sst{(7)}}}

\newcommand{\be}{\begin{equation}}
\newcommand{\ee}{\end{equation}}
\newcommand{\bs}{\begin{equation}\begin{split}}

\def\ba{\begin{array}}
\def\ea{\end{array}}
\newcommand{\g}{\gamma}
\newcommand{\ghat}{\hat{\gamma}}

\newcommand{\R}{\mathds{R}}
\newcommand{\C}{\mathds{C}}
\newcommand{\Q}{\mathds{H}}
\newcommand{\Oct}{\mathds{O}}
\newcommand{\Al}{\mathds{A}}
\newcommand{\id}{\mathds{1}}

\newcommand{\gamo}[3]{\Gamma^{#1}_{{#2}{#3}}}
\newcommand{\bgamo}[3]{\bar\Gamma^{#1}_{{#2}{#3}}}

\newcolumntype{L}[1]{>{\raggedright\let\newline\\\arraybackslash\hspace{0pt}}m{#1}}
\newcolumntype{C}[1]{>{\centering\let\newline\\\arraybackslash\hspace{0pt}}m{#1}}
\newcolumntype{R}[1]{>{\raggedleft\let\newline\\\arraybackslash\hspace{0pt}}m{#1}}

\DeclareMathOperator{\SO}{SO}

\DeclareMathOperator{\End}{End}

\newcommand{\ehat}{\hat{e}}

\newcommand{\bea}{\begin{eqnarray}}
\newcommand{\eea}{\end{eqnarray}}

\newcommand{\susy}{\mathcal{N}}
\newcommand{\N}{\mathcal{N}}

\begin{document}

\title{An octonionic formulation of the M-theory algebra}
\author{A. Anastasiou}
\email[]{alexandros.anastasiou07@imperial.ac.uk}
\author{L. Borsten}
\email[]{leron.borsten@imperial.ac.uk}
\author{M. J. Duff}
\email[]{m.duff@imperial.ac.uk}
\author{L. J. Hughes}
\email[]{leo.hughes07@imperial.ac.uk}
\author{S. Nagy}
\email[]{s.nagy11@imperial.ac.uk}

\affiliation{Theoretical Physics, Blackett Laboratory, Imperial College London, London SW7 2AZ, United Kingdom}

\date{\today}

\begin{abstract}
We give an octonionic formulation of the $\susy=1$ supersymmetry algebra in $D=11$, including all brane charges. We write this in terms of a novel outer product, which takes a pair of elements of the division algebra $\Al$ and returns a real linear operator on $\Al$. More generally, with this product comes the power to rewrite any linear operation on $\R^n$ ($n=1,2,4,8$) in terms of multiplication in the $n$-dimensional division algebra $\Al$. Finally, we consider the reinterpretation of the $D=11$ supersymmetry algebra as an octonionic algebra in $D=4$ and the truncation to division subalgebras.

\end{abstract}



\preprint{Imperial/TP/2014/mjd/01}

\maketitle

\section{Introduction}

A recurring theme in the study of supersymmetry and string theory is the connection to the four division algebras: the real numbers $\R$, the complex numbers $\C$, the quaternions $\Q$ and the octonions $\Oct$. See, for example, \cite{Kugo:1982bn,Duff:1987qa,Manogue:1993ja,Baez:2009xt,Baez:2010ye,Evans:1994cn,Borsten:2008wd,Anastasiou:2013cya}. The octonions are of particular interest in this context  since they may be used to describe representations of the Lorentz group in spacetime dimensions $D=10,11$, where string and M-theory live. Furthermore, the octonions provide a natural explanation \cite{Borsten:2013bp,Anastasiou:2013hba} for the appearance of exceptional groups as the U-dualities of supergravities \cite{Cremmer197848} and M-theory \cite{Duff:1990hn,Hull:1994ys}.

A $D=11$ spinor with 32 components may be packaged as a 4-component octonionic column vector \cite{Baez:2010ye,Toppan:2003ry}. This has prompted the question of how to write the algebra of $D=11$ supergravity (or `M-algebra') using octonionic supercharges $Q$. This was explored in \cite{Toppan:2003ry} where the problem was highlighted that the apparently natural choice of octonionic matrices could not provide enough degrees of freedom to account for all of M-theory's brane charges. Another fundamental question that arises when writing the $\{Q,Q\}$ algebra in this way is whether or not the usual anti-commutator is really the appropriate object to study, given that the fermionic supercharges are written over a non-commutative and non-associative algebra $\Oct$.

In the present paper we tackle this problem by introducing a novel outer product, which takes a pair of elements belonging to a division algebra $\Al$ and returns a real linear operator on $\Al$, expressed using multiplication in $\Al$. This product enables one to rewrite any expression involving $n\times n$ matrices and $n$-dimensional vectors in terms of multiplication in the $n$-dimensional division algebra $\Al$. We solve the problem of the octonionic M-algebra using this product, which allows a derivation of the correct $\{Q,Q\}$ bracket. In the final section we consider ``Cayley-Dickson halving'' the octonionic M-algebra, which corresponds to its reinterpretation as the maximal supergravity algebra in $D=7,5,4$. For example, the M-algebra may be considered to be an octonionic rewriting of the $D=4$, $\mathcal{N}=8$ supersymmetry algebra; from this perspective the $D=4$, $\mathcal{N}=1$ algebra comes from a truncation $\Oct\rightarrow\R$.

\section{The Division Algebras}

A normed division algebra is an algebra $\Al$
equipped with a positive-definite norm satisfying the condition 
\be
|\hspace{-0.2mm}|xy|\hspace{-0.2mm}|=|\hspace{-0.2mm}|x|\hspace{-0.2mm}|\hspace{0.4mm}|\hspace{-0.2mm}|y|\hspace{-0.2mm}|.
\ee
Remarkably, there are only four such algebras: $\R$, $\C$, $\Q$ and $\Oct$, with dimensions $n=1,2,4$ and $8$, respectively.

A division algebra element $x\in\Al$ is written as the linear combination of $n$ basis elements with real coefficients: $x=x_{a}e_{a}$, with $x_a\in\R$ and $a=0,\cdots,(n-1)$. One basis element $e_0=1$ is real; the other $(n-1)$ $e_i$ are imaginary:
\be
e_0^2=1,~~~~e_i^2=-1,
\ee
where $i=1,\cdots,(n-1)$. In analogy with the complex case, we define a conjugation operation indicated by *, which changes the sign of the imaginary basis elements:
\be
{e_0}^*=e_0,~~~~{e_i}^*=-e_i.
\ee
The multiplication rule for the basis elements of a division algebra is given by:
\be\label{OCTMULT}
\begin{split}
&{e_a}{e_b}=\left(\delta_{a0}\delta_{bc}+\delta_{0b}\delta_{ac}-\delta_{ab}\delta_{0c}+C_{abc}\right)e_c\equiv\gamo{a}{b}{c}e_c,\\
&{e_a^*}{e_b}=\left(\delta_{a0}\delta_{bc}-\delta_{0b}\delta_{ac}+\delta_{ab}\delta_{0c}-C_{abc}\right)e_c\equiv\bgamo{a}{b}{c}e_c,
\end{split}
\ee
where we define the structure constants\footnote{The unusual choice of index structure is for later convenience - see equations (\ref{CLIFFORD}) and (\ref{PSIMULT}).}
\begin{eqnarray}\label{STCON}
\Gamma^a_{bc}&=&\delta_{a0}\delta_{bc}+\delta_{b0}\delta_{ac}-\delta_{ab}\delta_{c0}+C_{abc},\\
\bar\Gamma^a_{bc}&=&\delta_{a0}\delta_{bc}-\delta_{b0}\delta_{ac}+\delta_{ab}\delta_{c0}-C_{abc}~~\Rightarrow~~\Gamma^a_{bc}=\bar\Gamma^a_{cb}.\nonumber
\end{eqnarray}
The tensor $C_{abc}$ is totally antisymmetric with $C_{0ab}=0$, so it is identically zero for $\Al=\R,\C$. For the quaternions $C_{ijk}$ is simply the permutation symbol $\varepsilon_{ijk}$, while for the octonions the non-zero $C_{ijk}$ are specified by the set of oriented lines of the Fano plane, see \cite{Baez:2001dm}.

One of the most important properties of the division algebras is that they provide a representation of the $\SO(n)$ Clifford algebra. This is reflected in the structure constants, which satisfy
\bs\label{CLIFFORD}
&\Gamma^a\bar\Gamma^b+\Gamma^b\bar\Gamma^a=2\delta^{ab}\id,\\
&\bar\Gamma^a\Gamma^b+\bar\Gamma^b\Gamma^a=2\delta^{ab}\id.
\end{split}\ee
In other words, we have the interpretation that multiplying a divison algebra element $\psi$ by the basis element $e_a$ has the effect of multiplying $\psi$'s components by the gamma matrix $\bar\Gamma^a$:
\be\label{PSIMULT}
e_a\psi=e_a e_b \psi_b= \Gamma^a_{bc}e_c \psi_b = e_c\bar\Gamma^a_{cb}\psi_b.
\ee
This property is essential for many of the applications of division algebras to physics, including that of this paper.

A natural inner product \cite{Baez:2001dm} on $\Al$ is given by:
\be
\langle{x}|{y}\rangle=\frac{1}{2}\left(x^*y+y^*x\right)=x_ay_a\hspace{0.4cm}\text{i.e.}\hspace{0.4cm}\langle{e_a}|{e_b}\rangle=\delta_{ab}.
\ee
This is just the canonical inner product on $\R^n$. 

\section{A New Outer Product}

It is interesting to see what other linear operations on $\R^n$ look like when written in terms of the division-algebraic multiplication rule. This was explored in \cite{DeLeo:1996kg}, but we take a different approach here. Consider the following general problem. Given some linear operator on $\R^n$ expressed as an $n\times n$ matrix $M_{ab}$, we would like to find an operator $\hat{M}$ on the division algebra $\Al$ such that $\hat{M}$ has the effect of multiplying the components of $x=x_ae_a\in\Al$ by $M_{ab}$:
\be\label{Mhat}
\hat{M}x\equiv e_aM_{ab}x_b.
\ee
An explicit form for this operator can be found using the inner product above. First we rewrite
\be
\begin{split}
M_{ab}&=M_{cd}\langle{e_a}|{e_c}\rangle\langle{e_b}|{e_d}\rangle\\
&=\frac{1}{2}M_{cd}\big\langle e_a | e_c(e_d^*e_b)+e_c(e_b^*e_d)\big\rangle.
\end{split}
\ee
Now it is clear that the operator
\be
\hat{M}\equiv\frac{1}{2}M_{cd}\Big(e_c\big(e_d^*\cdot\big)+e_c\big((\cdot)^*e_d\big)\Big),
\ee
where a dot represents a slot for an octonion, has matrix elements
\be
\langle e_a | \hat{M}e_b\rangle = M_{ab}.
\ee
This suggests that we write the outer product for division algebra elements using their multiplication rule, defining:
\begin{eqnarray}
\times~:\Al&\otimes&\Al\rightarrow \End(\Al)\label{OUTER}\\
e_a&\otimes& e_b \mapsto e_a\times e_b \equiv \frac{1}{2}\Big(e_a\big(e_b^*\cdot\big)+e_a\big((\cdot)^*e_b\big)\Big).\nonumber
\end{eqnarray}
With the new product comes the power to rewrite any expression involving $n\times n$ matrices and $n$-dimensional vectors in terms of multiplication in the $n$-dimensional division algebra $\Al$.

It is useful to note various equivalent ways of writing the outer product above:
\bs\label{VERSIONS}
e_a\times e_b&=\frac{1}{2}\Big(e_a\big(e_b^*\cdot\big)+e_a\big((\cdot)^*e_b\big)\Big)\\
&=\frac{1}{2}\Big(\big(\cdot e_b^*\big)e_a+\big(e_b(\cdot)^*\big)e_a\Big)\\
&=\frac{1}{2}\Big(e_a\big(e_b(\cdot)^*\big)+e_a\big(\cdot e_b^*\big)\Big)\\
&=\frac{1}{2}\Big(\big((\cdot)^*e_b\big)e_a+\big(e_b^*\cdot \big)e_a\Big).
\end{split}\ee
Due to the alternativity of the division algebras we also have
\be
e_a\big(e_b^*\cdot\big)+e_a\big((\cdot)^*e_b\big)
=\big(e_ae_b^*\big)(\cdot)+\big(e_a(\cdot)^*\big)e_b,
\ee
and similarly for the other four possibilities above.

\section{Octonionic Spinors in $D=11$}

In $D=11$ the Majorana spinor may be written as a 32-component real column vector. However, if we consider $\R^{32}$ as the tensor product $\R^4\otimes\R^8\cong\R^4\otimes\Oct$ then we can write this as a 4-component octonionic column vector
\be\label{SPINOR}
\lambda = \begin{pmatrix} \lambda_1 \\ \lambda_2 \\ \lambda _3 \\ \lambda_4 \end{pmatrix},~~~~ \lambda_\alpha \in \Oct, ~~\alpha =1,2,3,4.
\ee
A natural set of generators $\{\gamma^M\}=\{\gamma^0,\gamma^{a+1},\gamma^9,\gamma^{10}\}$, $M=0,1,\ldots,10$ for the $4\times4$ octonionic Clifford algebra is then given by
\be\label{GAMMAS}
\begin{split}
\gamma^0&=\begin{pmatrix}0&0 & 1&0\\ 0&0 & 0& 1\\-1&0 & 0&0\\0&-1 & 0&0\end{pmatrix},~~~\gamma^{a+1}=\begin{pmatrix}0&0 & 0&e_a^*\\ 0&0 & e_a& 0\\0&e_a^* & 0&0\\e_a&0 & 0&0\end{pmatrix},\\
\gamma^9&=\begin{pmatrix}0&0 & 1&0\\ 0&0 & 0& -1\\1&0 & 0&0\\0&-1 & 0&0\end{pmatrix},~~~~\gamma^{10}~=\begin{pmatrix}1~&0 & 0&0\\ 0~&1 & 0& 0\\0~&0 & -1&0\\0~&0 & 0&-1\end{pmatrix},
\end{split}
\ee
with $a=0,1,\ldots,7$. These matrices satisfy
\be
\gamma^M\gamma^N+\gamma^N\gamma^M=2\eta^{MN}\id,
\ee
and the infinitesimal Lorentz transformation of the spinor $\lambda$ is
\be
\delta\lambda=\frac{1}{4}\omega_{MN}\gamma^M(\gamma^N\lambda),
\ee
where $\omega_{MN}=-\omega_{NM}$. In general, the action of the rank $r$ Clifford algebra element on $\lambda$ can be written
\be\label{rankr}
\g^{[M_1}\Big(\g^{M_2}\big(\ldots(\g^{M_{r-1}}(\g^{M_r]}\lambda))\ldots\big)\Big).
\ee
The positioning of the brackets in the above expression follows from repeated application of (\ref{PSIMULT}); non-associativity matters only for the imaginary gamma matrices $\g^{i+1}$, which provide a representation of the SO(7) Clifford algebra. If we define an operator $\ghat^M$, whose action is left-multiplication by $\g^M$, then we can think of the rank $r$ Clifford algebra element as the operator
\be
\ghat^{[M_1M_2\ldots M_r]}\equiv\ghat^{[M_1}\ghat^{M_2}\ldots\ghat^{M_r]},
\ee
where the operators $\ghat^M$ must be composed as
\be
\ghat^M\ghat^N\lambda=\g^M(\g^N\lambda)\neq (\g^M\g^N)\lambda.
\ee
This ensures that the action of $\ghat^{[M_1M_2\ldots M_r]}$ on a spinor is given by (\ref{rankr}), as required.

\section{The Octonionic M-Algebra}

The anti-commutator of two supercharges in the $D=11$ supergravity theory is conventionally written as the `M-algebra' \cite{vanHolten:1982mx,Townsend:1995gp}
\be\begin{split}\label{CONVENTIONAL}
\{Q_{\bar\alpha},Q_{\bar\beta}\}=&(\g^MC)_{\bar\alpha\bar\beta}P_M+(\g^{MN}C)_{\bar\alpha\bar\beta}Z_{MN}\\
&+(\g^{MNPQR}C)_{\bar\alpha\bar\beta}Z_{MNPQR},
\end{split}
\ee
where $\bar\alpha,\bar\beta=1,\ldots,32$, $P_M$ is the generator of translations and $Z_{MN}$ and $Z_{MNPQR}$ are the brane charges. The charge conjugation matrix $C_{\bar\alpha\bar\beta}$ serves to lower an index on each of the gamma matrices.

The left-hand side is a symmetric $32\times32$ matrix with 528 components, while the terms on the right-hand side consist of the rank 1, 2 and 5 Clifford algebra elements, which form a basis for such symmetric matrices. In terms of $\SO(1,10)$ representations:
\be
(\bold{32}\times\bold{32)_{\text{\footnotesize{Sym}}}=11+55+462}.
\ee
We would like to write this algebra in terms of $4\times4$ octonionic matrices. However, the space of octonionic $4\times4$ matrices is of dimension $16\times 8= 128$, and hence naively does not carry nearly enough degrees of freedom to write (\ref{CONVENTIONAL}).

The solution to this problem is to use the octonionic Clifford algebra operators $\ghat^{[M_1M_2\ldots M_r]}$ defined in the previous section. These operators (including all ranks $r$) span a space of dimension $32\times32=1024$. In other words, their octonionic matrix elements are
\be
\langle e_a|\ghat^M{}_{\alpha}{}^{\beta} e_b\rangle=\gamma^M{}_{\alpha a}{}^{\beta b},~~~\alpha,\beta =1,2,3,4,
\ee
and if we think of $\alpha a$ as a composite spinor index $\bar\alpha=1,\ldots,32$, then the set of $\{\gamma^M{}_{\bar\alpha}{}^{\bar\beta}\}$ generates the usual real Clifford algebra as in (\ref{CONVENTIONAL}).

For the charge conjugation matrix, we define the $4\times 4$ real matrix (which is numerically equal to $\g^0$ but with a different index structure)
\be
C_{\alpha\beta}=\begin{pmatrix}0&0 & 1&0\\ 0&0 & 0& 1\\-1&0 & 0&0\\0&-1 & 0&0\end{pmatrix}.
\ee
The octonionic matrix elements of this are then trivially
\be
C_{\alpha a\beta b}=\langle e_a|C_{\alpha\beta} e_b\rangle=C_{\alpha\beta}\delta_{ab},
\ee
which can be identified with the $32\times 32$ matrix:
\be
C_{\bar\alpha\bar\beta}=C_{\alpha a\beta b}=C_{\alpha\beta}\delta_{ab}.
\ee

Armed with these tools, the right-hand side can then be written over $\Oct$ simply by replacing $\bar\alpha\rightarrow\alpha$ and putting hats on the gammas:
\be\begin{split}\label{RHS}
&(\ghat^MC)_{\alpha\beta}P_M+(\ghat^{MN}C)_{\alpha\beta}Z_{MN}\\
&+(\ghat^{MNPQR}C)_{\alpha\beta}Z_{MNPQR}.
\end{split}
\ee

With the identification $\bar\alpha=\alpha a$ we can also write the left-hand side of (\ref{CONVENTIONAL}) in terms of the composite indices:
\be\label{COMPQ}
\{Q_{\bar\alpha},Q_{\bar\beta}\}=\{Q_{\alpha a},Q_{\beta b}\}.
\ee
Now, the expression (\ref{RHS}) is an octonionic operator with matrix elements as on the right-hand side of (\ref{CONVENTIONAL}), so on the left we require an octonionic operator
\be
\widehat{\{Q_{\alpha},Q_{\beta}\}}
\ee
with matrix elements given by (\ref{COMPQ}). The required operator is obtained simply by contracting (\ref{COMPQ}) with the outer product $e_a\times e_b$ defined in (\ref{OUTER}):
\be
\widehat{\{Q_{\alpha},Q_{\beta}\}}\equiv\{Q_{\alpha a},Q_{\beta b}\}e_a\times e_b.
\ee
The octonionic formulation of the M-algebra is then
\bs\label{FINAL}
\widehat{\{Q_{\alpha},Q_{\beta}\}}=&(\ghat^MC)_{\alpha\beta}P_M+(\ghat^{MN}C)_{\alpha\beta}Z_{MN}\\
&+(\ghat^{MNPQR}C)_{\alpha\beta}Z_{MNPQR}.
\end{split}\ee
Using the first two versions of the outer product given in (\ref{VERSIONS}), we could write the left-hand side as
\bs\label{LHS}
\widehat{\{Q_{\alpha},Q_{\beta}\}}~=~~\frac{1}{2}\Big(&\big(Q_\alpha Q_\beta^*\big)(\cdot)+(\cdot)\big( Q_\beta^* Q_\alpha\big)\\+&\big(Q_\alpha(\cdot)^*\big)Q_\beta+Q_\beta\big((\cdot)^*Q_\alpha\big)\Big).
\end{split}\ee
The first two terms look similar to the more intuitive anti-commutator $\{Q_\alpha,Q_\beta^*\}$, explored in \cite{Toppan:2003ry}, but to reproduce the full M-algebra we require all four terms above.

\section{Relation To Lower Dimensions}

It is interesting to consider the octonionic version of the supersymmetry algebra after an $\bold{11=4+7}$ split:
\be
\SO(1,10)\supset \SO(1,3)\times\SO(7).
\ee
Seven of the Clifford algebra generators $\g^{i+1}$ are imaginary, while the other four are real. This suggests that we split the dimensions as follows:
\bs
M=0,1,\ldots,10 ~~\rightarrow~~i+1&=1,\ldots,8,\\ \mu&=0,1,9,10.
\end{split}\ee

In $D=4$ we regard the $D=11$ octonionic spinor $Q_{\alpha a}e_a$ as eight 4-component Majorana spinors $Q_{\alpha a}$, which we may leave packaged as an `internal' octonion. This transforms as the spinor $\bold{8}$ of SO(7). The $D=4$ interpretation of the octonionic gamma matrices  is as follows:
\be
\hat{\g}_{i+1}=\g_* \hat{e}_i,
\ee
where $\hat{e}_i$ denotes the operator whose action is left-multiplication by $e_i$ and $\g_*$ (otherwise known as $\g_5$) is the highest rank Clifford element:
\be
\g_*=-\g^0\g^1\g^9\g^{10}=\begin{pmatrix}0&0 & 0&-1\\ 0&0 & 1& 0\\0&-1& 0&0\\1&0 & 0&0\end{pmatrix}.
\ee
The matrix $C_{\alpha\beta}$ is just the charge conjugation matrix in $D=4$.

We do not split the $M,N$ indices of equation (\ref{FINAL}) into $\mu$ and $i$ parts here, as the expression of the right-hand side itself is not particularly illuminating. The result is a copy of the $\N=8$ supersymmetry algebra written over the octonions. The interesting point is that the $D=11$ supersymmetry algebra \emph{can} be reinterpreted as an octonionic $D=4$ algebra.

\enlargethispage{21.8cm}

More generally, the spinor and associated gamma matrices defined in (\ref{SPINOR}) and (\ref{GAMMAS}) correspond to those of $D=4,5,7$ if we replace $\Oct$ with $\R,\C,\Q$, respectively - see Table \ref{ALGEBRAS}. This means that in this framework the minimal supersymmetry algebra in these dimensions is written over $\R,\C,\Q$, while doubling the amount of supersymmetry corresponds to Cayley-Dickson doubling the division algebra. This process terminates when we reach maximal supersymmetry, i.e. when the Cayley-Dickson process takes us to $\Oct$, the largest normed division algebra.

\begin{table}[h!]\vspace{0.5cm}
\begin{tabular}{C{1cm}|C{1.5cm}C{1.5cm}C{1.5cm}C{1.5cm}}

\hline\hline
$D\hspace{0.1cm}\backslash\hspace{0.1cm}\mathcal{N}$
& $1$ & $2$ & $4$ & $8$ \\ 
\hline
\\[-0.3cm]
$11$  & $\Oct^4$ \\ 
$7$ & $\Q^4$ & $\Oct^4$
\\ 
$5$ & $\C^4$ & $\Q^4$
& $\Oct^4$ \\ 
$4$ & $\R^4$ & $\C^4$ & $\Q^4$ & $\Oct^4$
\\ 
\hline\hline
\end{tabular}
\caption{\footnotesize{A summary of the division algebraic parameterisation of spinors used in $D=n+3$ supersymmetry algebras. Note that supersymmetry algebras sharing the same $\Al$ are equivalent and that Cayley-Dickson doubling $\Al$ corresponds to doubling $\N$, or equivalently climbing upwards in dimension $D$.}}\label{ALGEBRAS}
\end{table}

\newpage
\begin{table}[ht]\vspace{0.5cm}
\begin{tabular}{C{1cm}|C{1.5cm}C{1.5cm}C{1.5cm}C{1.5cm}}

\hline\hline
${D}\hspace{0.1cm}\backslash\hspace{0.1cm}\mathcal{N}$
& $1$ & $2$ & $4$ & $8$ \\ 
\hline
\\[-0.3cm]
$11$ & $~\R^{32}$ \\ 
$7$ & $~\R^{16}$ & $~\C^{16}$\\ 
$5$ & $\R^8$ & $\C^8$& $\Q^8$ \\ 
$4$ & $\R^4$ & $\C^4$ & $\Q^4$ & $\Oct^4$
\\ 
\hline\hline
\end{tabular}
\caption{\footnotesize{An alternative parameterisation of spinors used in $D=n+3$ supersymmetry algebras. From this point of view the octonions single out $D=4$.}}\label{ALGEBRAS2}
\end{table}

The above discussion serves to emphasise the correspondence between the octonions and maximal supersymmetry in various dimensions. Rather than  thinking of the M-theory algebra as an eleven-dimensional real algebra, it may be fruitful to think of it as a four-dimensional octonionic one, as in Table \ref{ALGEBRAS2}.

\begin{acknowledgements}

The work of LB is supported by an Imperial College Junior Research Fellowship. The work of MJD is supported by the STFC under rolling grant ST/G000743/1.

\end{acknowledgements}

\bibliography{Octonions}

\end{document}